# Sentiment Analysis and Sarcasm Detection of Indian General Election Tweets

Arpit Khare, Amisha Gangwar, Sudhakar Singh*[0000-0002-0710-924X], Shiv Prakash[0000-0002-4240-0390]

Department of Electronics and Communication, University of Allahabad, Prayagraj, India
{ arpitkhare33, amishagangwar21}@gmail.com, {sudhakar,
shivprakash}@allduniv.ac.in
*Corresponding Author

**Abstract.** Social Media usage has increased to an all-time high level in today's digital world. The majority of the population uses social media tools (like Twitter, Facebook, YouTube, etc.) to share their thoughts and experiences with the community. Analysing the sentiments and opinions of the common public is very important for both the government and the business people. This is the reason behind the activeness of many media agencies during the election time for performing various kinds of opinion polls. In this paper, we have worked towards analysing the sentiments of the people of India during the Lok Sabha election of 2019 using the Twitter data of that duration. We have built an automatic tweet analyser using the Transfer Learning technique to handle the unsupervised nature of this problem. We have used the Linear Support Vector Classifiers method in our Machine Learning model, also, the Term Frequency Inverse Document Frequency (TF-IDF) methodology for handling the textual data of tweets. Further, we have increased the capability of the model to address the sarcastic tweets posted by some of the users, which has not been yet considered by the researchers in this domain.

## 1 Introduction

In the digital era of modern times, everything is shifting towards online mode, and in the pandemic time of COVID-19, this shift towards online mode has increased drastically. Let it be, studies, work, doctors' practice, banking, shopping, fitness, surveys, etc. everything has started working online. In today's world, to share our opinion with other people, we use online social networking platforms like Twitter, Facebook, YouTube, etc. Very often, people use these platforms to share their experiences about some incidents or some issues they are facing. The usage of these platforms allows the user to increase the range of audience to which they can share their thoughts, as compared to the offline method for opinion sharing. The analysis of the opinion of the common public is very important for a wide range of organisations, let it be a political party, businessman, investors, and working professionals, everyone tries to understand the opinion and trends getting followed by the public.

When it comes to politics and elections, many parties try to recognise the trend of opinions getting followed among the citizens. Even news channels and different media houses, perform various kinds of opinion polls and survey polls to understand the opinions and sentiments of the common person. But, in the pandemic time of COVID-19, reaching out to someone physically is not possible at sometimes [1]. Because elections are a periodic activity that will occur regularly, even during the pandemic time, so to reduce the risk of the field workers of the polling agency for performing survey polls, sentiment analysis of political tweets is very significant. In this paper, we have used the political tweets of the Lok Sabha Elections – 2019 for getting the political sentiments of the people by performing the sentiment analysis techniques along with handling the sarcastic tweets which is not yet considered in the state-of-the-art works in this domain. Training and testing are done on Election – 2019 data where tweets are taken from the data science platform 'Kaggle' [2]. The paper presents the analysis of the tweets of the major political parties participating in the elections. It applied the Transfer learning approach for handling the unsupervised nature of the problem. The main contributions of this paper are as follows.

- Developing a robust sentiment analysis model for election data.
- Making the model capable enough to handle the sarcastic tweets.
- Implementing the transfer learning approach for handling the unsupervised nature of the problem.
- Analysing the results of the model with respect to the actual election results.

The paper is structured section-wise as follows. Section 1 introduces the paper. Section 2 presents the literature review, in which the related works in the domain of sentiment analysis and election results predictions are reviewed. Section 3 contains material and methods including the approaches proposed for handling the sentiment analysis and sarcasm detection of textual data. Section 4 is the experimental evaluation of the sentiment analysis and sarcasm detection model. Finally, Section 6 concludes the paper with the scope of future work.

## 2    Related Works

Various researchers have been doing Sentiment analysis of Twitter data in various application fields from past times which is not limited to only political tweets. Pang and Lee [3] in 2008, developed an algorithm to mine the opinions of different people and analyse types of words. Jhanwar and Das [4] proposed a technique for studying the sentiments of the people of India by analysing the texts written in mixed languages of Hindi-English. Go et al. [5] tried to train a sentiment analysis algorithm to detect whether a tweet is a positive tweet about any specific subject by using emoticons. In the same domain, Pak and Paroubek [6] tried to expand the functionality by means of subjectively and objectively of different words which can create classifiers. These classifiers can be used to collect data to regulate a tweet's sentiment. In [7], an open-source application called VADER [7] was introduced. This is a rule-based application for finding the sentiment scores for textual data. Zhang et al. [8] examined the COVID-19 tweets in four cities of Canada and four cities of the USA. Sentiment intensity scores are examined through VADER [7] and NRC [9] methods and visualized the data during the pandemic. Das and Bandyopadhyay [10] tried many approaches for concluding the sentiments behind various tweets. They used the approaches of interactive game, bilingual dictionary, and WordNet library. In the literature [11] and [12], the authors have used a dictionary-based approach in which they created a small set of opinion words and then searched in a large collection of texts for growing the dictionary, in the same way as is done in WordNet [13]. Taboada et al. [14] also introduced a sentiment analysis method using the lexicon-based approach. The authors used a dictionary of positive or negative visualized words to analyse the tweets. Bhadane et al. [15] reviewed various methods used to perform analysis on some natural language textual data, as per the sentiments expressed in that data, i.e. whether the text is of positive emotion or negative emotion.

Political tweets have been a center of attraction for many stakeholders. A number of authors have contributed towards sentiment analysis on election and political tweets. Hamling and Agrawal [16] have analysed the 2016 US election related tweets. These tweets were used to find a correlation between the tweet sentiments and the election results. They used the Sentiwordnet library [17] for finding out the sentiment scores. Values for positive words were ranged from +0.0625 to +1.0; and -0.0625 to -1.0 for negative values. For instance, positive keyword like "helpful" would receive a value of +0.125 from the sentiment algorithm, or a strongly negative word like "unhappy" would receive a value of -0.25. In their algorithm, they were not able to handle the sarcastic tweets done by some of the users, since sarcasm is also a general form of expression of opinion, so the importance of sarcasm detection can't be overlooked in this domain. In [18], the authors tried to extract the positive and negative sentiments and emotions of the people for the common political parties by computing a distance measure. It denotes the closeness of various tweets. More closely the tweets are for different political parties, it indicates the chances of a tough and close fight among the different parties.

What people think is directly related to what they post over social media, many investigations have proved to predict election results by performing sentiment analysis of Twitter data, such as, using the lexicon method for the Swedish elections [19]. Jose and Chooralil [20] have implemented a new method by using Twitter's streaming API for the data collection process. They tried to extract the sentiments and information from the tweets by using lexical tools like WordNet [13] and SentiWordNet [10]. Also, to increase their efficiency, they used a method for handling negation in the data pre-processing stage.

The usage of Twitter by politicians and their political campaigns has been a subject of interest for researchers. In the 2008 USA elections, the campaign routines of Barack Obama have increased the interest in the role of Twitter in political battles [21][22]. In [23] and [24], it is reported that many US Congress members in their Twitter venture posted their opinions about political issues on Twitter and the issues related to their election area.

Sharmistha Chatterjee [25] used crawling twitter data through API for performing the sentiment analysis of the two major parties BJP and INC. She used standard ML and Deep learning algorithms for mood classification of the two major parties. She crawled and merged the tweets every week for some months. The WordCloud and N-gram Model [26] was used for Sentiment Representation. She also added an additional location mapping feature for the tweets and used the retweet frequency distribution. Sharma and Moh [27] attempted to predict Indian election results using sentiment analysis on Hindi Twitter data. They fetched a total of 42,345 tweets in Hindi language. Then they performed data cleaning to remove irrelevant tweets and left with toal 36,465 tweets, and after that labeled the data manually on these 36,465 tweets, making the unsupervised problem a supervised one. Then they used I Bayes, support vector machine on the Twitter data.

Gaikar and Sapare [28] also predicted the results of General Elections-2019 in India using the LSTM Neural Network approach. They used over 1500 labeled tweets, which were labeled with labels like positive, negative, and neutral for the training of their model. In real-time, they used the Twitter API to extract a total of 40,000 tweets from Jan 2019 to Mar 2019 related to elections to test their model. They also visualized their results using word clouds and compared them with ABP-C and India Today Survey results for the elections. Ansari et al. [29] also performed a classification of the tweets related to the General Elections-2019 in India, they employed the LSTM model to perform the classification process of the Twitter text. They used the classification model to predict the inclination of tweets to infer the results of elections. With respect to the Indian election, Sharma and Ghose [30] performed text mining using the named entity recognition to filter out the unrelated tweets. For performing sentiment analysis on the related tweets they used the model Rapid-Miner AYLIEN [31]. Naiknaware and Kawathekarm [32] used the Sentiment analysis score method of R programming language to execute the sentiment analysis on the election related tweets for General Election-2019 of India. Bose et al. [33] in the context of political tweets, used the NRC emotion lexicon approach for finding the overall tone of the event. Then they used the deep learning tool ParallelDots that can categorize the tweets into positive, negative, and neutral categories. A strategy known as Adaptive Neuro-Fuzzy Inference System (ANFIS) is proposed by Katta and Hegde [34]. The Fuzzy based ontology is made by implementing Non-Linear SVM classifier analysis to improve the fuzzy principles. They concluded that an ANFIS Non-linear SVM-based model for sentiment analysis of social media text is less complex and provides high accuracy. In [35], Bansal and Srivastava used a Lexicon-based approach for Twitter sentiment analysis for the vote share prediction using the emojis and n-gram features. Hitesh et al. [36] performed real-time sentiment analysis of the 2019 General elections in India using Word2vec and Random Forest Model. Joseph [37] used the Decision Tree classifier approach for predicting the outcomes for Indian general elections-2019. He considered the tweets of the English language. His approach was to perform the mood mapping of people over a timely basis during different phases of elections. Prediction of Indonesia's election results was performed by Kristiyanti et al. [38] using the Support Vector Machine (SVM) with selection features of Particle Swarm Optimisation (PSO) and Genetic Algorithm (GA). They tried to perform predictions for the post of President and Vice president of Indonesia. Hidayatullah et al. [39] also performed the prediction of Indonesia's election results using the deep learning approach using various algorithms like Convolutional Neural

Network (CNN), Long short-term memory (LSTM), CNN-LSTM, Gated Recurrent Unit (GRU)-LSTM, and Bidirectional LSTM. They compared the results with various traditional machine learning algorithms and concluded that the Bidirectional LSTM achieved the best accuracy.

The various methodologies have been introduced for opinion and sentiment mining, but all these are broadly classified into 2 major groups. One is a Machine learning approach and the second one is a Lexicon-based method, a linguistically-inclined method [40]. In this paper, we have used the Machine learning approach for opinion mining. There are two significant types of machine learning algorithms, supervised learning and unsupervised learning. The supervised learning approach requires labeled data for the particular domain targeted to build a machine learning model. In contrast, the unsupervised approach is used when there is no labeled data for training the model. Sentiment analysis of tweets lies in the category of unsupervised learning since we do not have any assigned labels to the 2019 Loksabha election-related tweets dataset. To solve this lack of having the desired labeled dataset, we have used the transfer learning approach. Transfer learning is a type of approach for solving unsupervised problems. In this approach, the model is trained using some other problems' dataset (labeled dataset), and then the trained model is used for predictions on some new dataset which is not having any class labels attached to it. Although several works have been done on sarcasm detection [41], none of them have been performed in the domain of political tweets, so, this paper contributes in that direction. It applies sarcasm detection in the domain of election related tweets.

## 3 Material and Methods

### 3.1 Dataset

The testing of our trained model is done using the **India Lok Sabha Elections-2019** tweets. During elections, people used to express their views, opinions, and experiences related to major political parties of that time i.e. BJP and INC. The complete dataset of the election-related tweets is available on the data science platform Kaggle [2]. The dataset consists of the following fields.

*Last_updated:* This column contains the information about the time stamp at which the particular tweet was last updated.

*Tweet_id:* The unique id which is assigned to every tweet.

*Created_at:* This column contains the timestamp at which the tweet was created.

*Full_text:* The column contains the complete tweet text on which we will perform the text analysis.

*Quote_count:* This column contains the frequency the current tweet was retweeted with a comment.

*Reply_count:* This column contains the frequency the current tweet was replied to or commented on by any user.

*Retweet_count:* This column contains the frequency the current tweet was retweeted.

*Favorite_count:* This column contains the frequency of likes the current tweet is having.

In this paper, we have considered only the column with textual data for the sentiment analysis process,i.e. 'full_text' column.

### 3.2 Machine Learning Algorithms

**TF-IDF.** Term Frequency – Inverse Document Frequency (TF-IDF) [42] is a matrix that provides the word frequency table in a document/ sentence. Term frequency (TF(t,d)) is the measure for the word occurring in a document(d), whereas Document Frequency(DF(t)) is counter for the number of documents the word is occurring. Inverse Document Frequency (IDF(t)) gives the relative weight for a word. If the word is occurring often, its IDF(t) measure will be low whereas the IDF(t) will be high for less occurring words. Hence, it can be mathematically defined as Eq. (1) [43][44].

$$IDF(t) = \frac{N}{DF(t)} \tag{1}$$

where, *N* is Number of documents, *DF(t)* is Document Frequency and *IDF(t)* is Inverse Document Frequency.

In Eq. (1), *N* can be a large value hence it will explode the value of *IDF(t)* or *DF(t)* can possibly be 0 during query time. Hence the log function is taken to control the former once and the latter one is resolved by adding 1. Then the new equation for IDF(t) will be as Eq. (2) [43][44].

$$IDF(t) = \log \frac{N}{DF(t)+1} \tag{2}$$

Finally from Eq. 2, *TF-IDF* can be mathematically expressed as Eq. (3) [43][44].

$$TF - IDF[t,d] = TF(t,d) * \log \frac{N}{DF(t)+1} \tag{3}$$

**Linear SVC.** Support Vector Machine (SVM), a machine learning algorithm that can perform classification, regression, and outliers identification. Linear Support Vector Classifier (Linear SVC) is a classifying algorithm that gives out the best fitting line or hyperplane depending upon the dimensions of the problem. Dimension refers to the number of features. SVC is chosen because it ignores all the outliners and only chooses the best hyperplane to distinguish between the classes. Fig. 1 shows a Linear SVC example in two-dimensional space [45].

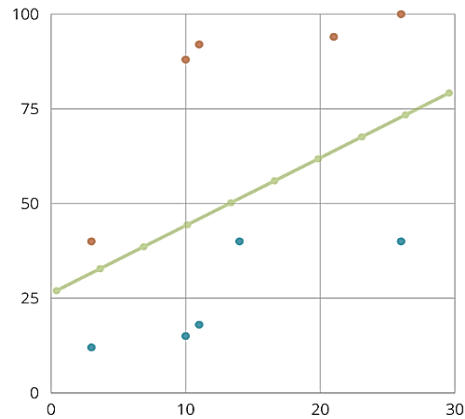

**Fig. 1.** Linear SVC example in 2-D space

### 3.3 Proposed Model

In this paper, we have tried to tackle the unsupervised nature of analysing sentiments of elections-2019 tweets by using transfer learning. We have used the standard Twitter review dataset available on Kaggle [2] for training our model using the Linear SVC and then performed the sentiment analysis on the actual data. Since the data was of textual format, we have used the TFIDF method for creating the Term Frequency Inverse Document Frequency matrix to deal with the textual data. This work improves over the study conducted on the US Presidential elections in which Sarcasm was not handled [16]. Fig. 2 and Fig. 3 illustrate the workflow of training of sentiment analysis and sarcasm detection model, and testing of the trained models on the election's dataset respectively.

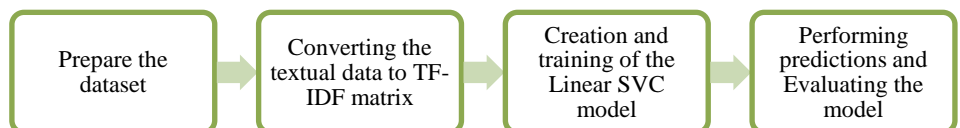

**Fig. 2.** The flow of work for the training of sentiment analysis and sarcasm detection model

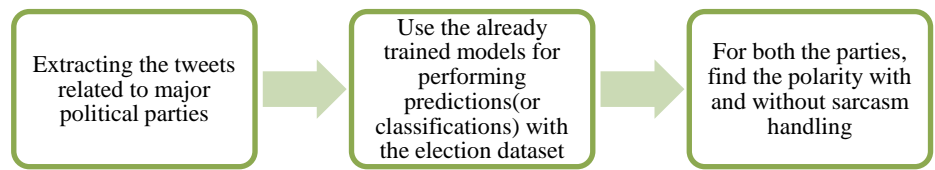

**Fig. 3.** The flow of work for the testing of the trained models on the election's dataset

Algorithms for training the sentiment analysis model, training of the sarcasm detection model, and transfer learning approach for sentiment analysis and sarcasm detection of the election's dataset are presented in 1, 2, and 3, respectively.

We have introduced an extra feature of sarcasm detection to the existing methodologies for performing sentiment analysis of political tweets. Since the nature of the incoming data makes the problem an unsupervised problem, i.e.it is not known beforehand which tweet is sarcastic and which one is not, we applied the Transfer learning approach to resolve this problem. For the training of the sarcasm detection model, a standard sarcasm detection dataset available on Kaggle is used, and then the trained model can be used to predict results on the actual tweets.

---

**Algorithm 1: Procedure for the training of the sentiment analysis model.**

1. Split the dataset into training and the testing data by the ratio of 70:30, setting the random state to be 42.
2. Use the TFIDFVectorizer of the sklearn library, transforming the textual data into the TFIDF matrix.
3. Create a pipeline passing the 2 phases for it as TFIDF and the LinearSVC model.
4. Fit the newly created pipeline with the training dataset along with the labels.
5. Perform the predictions using the testing data.
6. Compare the predicted results with the actual results by using the confusion_matrix and the classification_report which provides the accuracy score for the model.

---

**Algorithm 2: Procedure for the training of the sarcasm detection model.**

1. Use the TFIDFVectorizer of the sklearn library, transforming the textual data into the TFIDF matrix.
2. Create a pipeline passing the 2 phases for it as TFIDF and the LinearSVC model.
3. Fit the newly created pipeline with the training dataset along with the labels.
4. Perform the predictions using the testing data.
5. Compare the predicted results with the actual results using the confusion_matrix and the classification_report provides the accuracy score for the model.

---

**Algorithm 3: Procedure for using Transfer learning approach for sentiment analysis and sarcasm detection.**

1. Use the already created sentiment analysis and sarcasm detection models for performing predictions (or classifications) with the election dataset.
2. Add two new columns with the prediction results for both the models, i.e. one column having the sentiment result, whether positive or negative, while the second column holds the result for whether the tweet is sarcastic or not.
3. For both the parties, find the positive and negative popularity with and without sarcasm detection and display the results of both.
4. Plot the results for both the parties as bar graphs and pie charts.

## 4 Experiment Evaluation

### 4.1 System Configuration and Programing Environment

The experimentation is carried out on the Windows 10 Operating System installed on a machine having an Intel Core i5 processor, 8GB RAM, and a 2 TB hard disk drive. The implementation was done using the Programming Environment of Python 3.7 along with the Jupyter Notebook 6.4.3. The other adopted tools/libraries are as follows.

*Sklearn (Version 0.22):* Scikit-learn is a free software machine learning library in Python which provides various machine learning algorithms for Regressions, Classifications, and Clustering.

*Matplotlib(Version 3.4.3):* MatplotLib is a plotting library in Python which provides the API for plotting embedded plots for the data passed to it.

*Pandas(Version 1.3.2):* Pandas is a data table handling library in Python which provides the methods to perform all the operations on the data tables conveniently.

*Pipelines (Version 1.0):* Pipelines in python provide a sequence of transformation and prediction on the data as per the machine learning models passed to it.

*json library:* The json library can parse JSON from strings or files. The library parses JSON Python dictionary or list. It can also convert Python dictionaries or lists into JSON strings.

### 4.2 Results and Discussions

After the successful training of the model, to check the accuracy and classification quality of the model, we have used the confusion_matrix and classification_report methods available in the Python library sklearn. A confusion matrix is the tabular representation of the number of correct and incorrect classifications performed by the algorithms, whereas a classification report is a table that is used to find out the quality of the classifications performed by the classification algorithms. The results for the sentiment analysis model are as follows in Tables 1 and 2.

**Table 1.** Classification Report of Sentiment Analysis Model

| Type | Precision | Recall | F1-score | Support |
|---|---|---|---|---|
| Negative | 0.80 | 0.79 | 0.79 | 239819 |
| Positive | 0.79 | 0.80 | 0.80 | 240181 |
| Accuracy | | | 0.80 | 480000 |
| Macro Accuracy | 0.80 | 0.80 | 0.80 | 480000 |
| Weighted Average | 0.80 | 0.80 | 0.80 | 480000 |

**Table 2.** Classification Report of Sarcasm Detection Model

| Type | Precision | Recall | F1-score | Support |
|------|-----------|--------|----------|---------|
| 0 | 0.85 | 086 | 0.86 | 4498 |
| 1 | 0.82 | 0.81 | 0.82 | 3515 |
| Accuracy | | | 0.80 | 8013 |
| Macro Accuracy | 0.80 | 0.80 | 0.80 | 8013 |
| Weighted Average | 0.80 | 0.80 | 0.80 | 8013 |

Table 1 shows that our sentiment analysis model gets an accuracy of 80%, which is good to be used for transfer learning. Table 2 indicates that the sarcasm detection model achieves an accuracy of 84%, which makes it suitable for transfer learning.

We have used the above trained models for performing election tweets analysis for both without sarcasm handling and with sarcasm handling. The results for both the major parties BJP and INC for both the cases, i.e. without and with sarcasm are shown in Figs. 4 – 6 and Figs. 7 – 9 respectively.

The terms like positive tweets or positive polarity tweets indicate that the tweets are conveying some positive emotions about the particular political party. Similarly, on the other hand, terms like negative tweets or negative polarity tweets indicate that the tweets have some negative emotions/ sentiments about the political party.

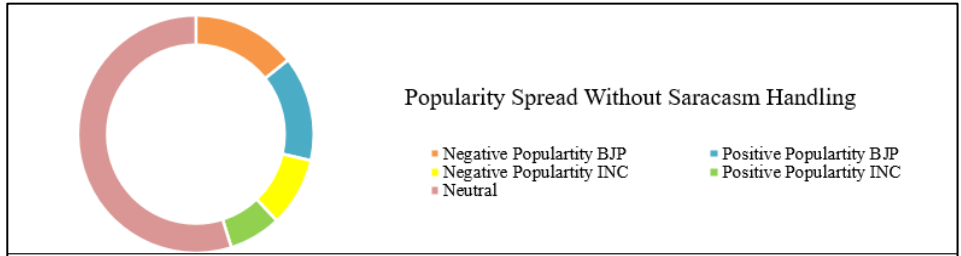

**Fig. 4.** Popularity spread of tweets without sarcasm handling

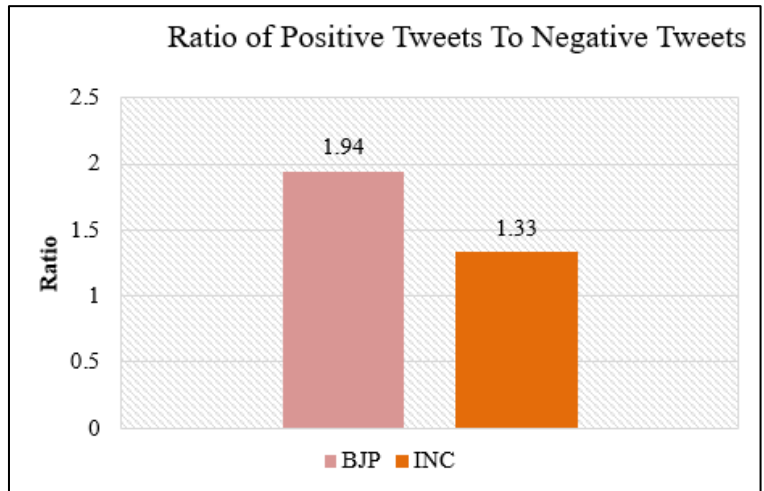

**Fig. 5.** Ratio of Positive to Negative tweets without sarcasm handling

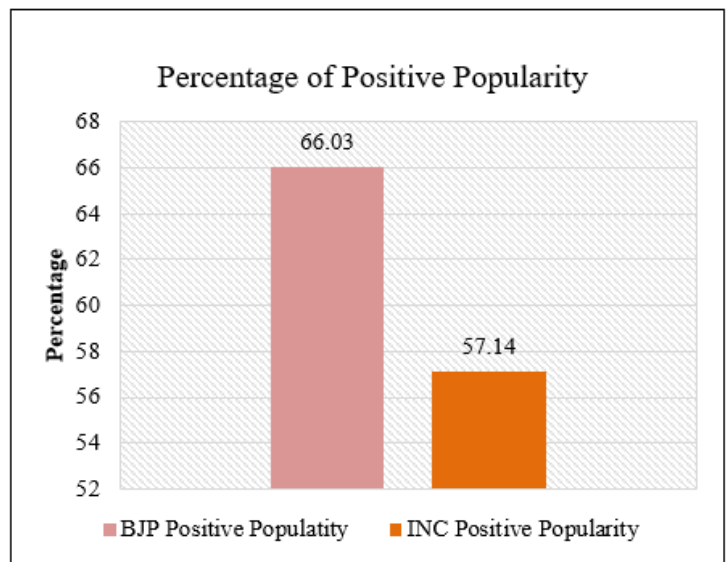

**Fig. 6.** Percentage of positive tweets among total tweets without sarcasm handling

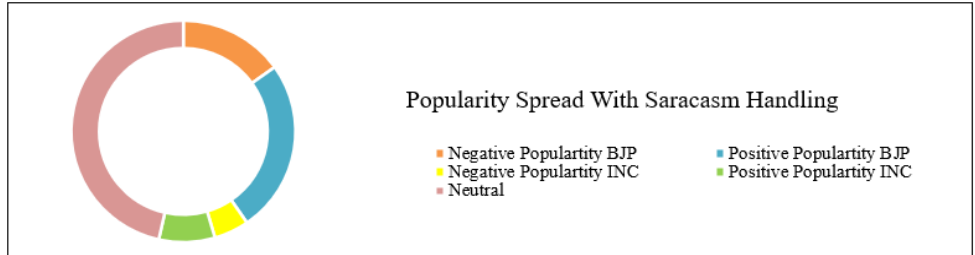

**Fig. 7.** Popularity spread of tweets with sarcasm handling

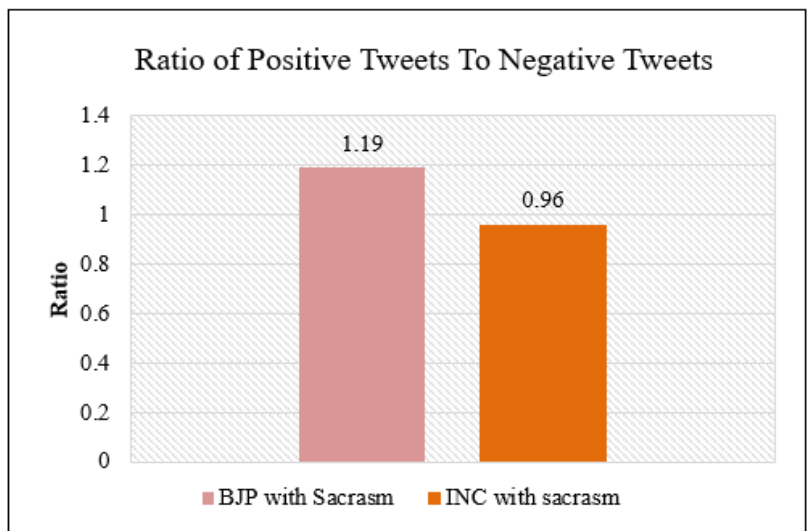

**Fig. 8.** The ratio of Positive to Negative tweets with sarcasm handling

Fig. 4 shows that 25.58% of the total tweets analysed without the sarcasm detection model were of positive polarity for the Bharatiya Janata Party (BJP). In the same manner, 13.16% of the total tweets were of negative polarity for the BJP. Whereas, Fig. 7 shows the results after handling sarcasm, i.e. 23.76% of the total tweets were actually of positive polarity for BJP, and 20.01% of the total tweets were of negative polarity for BJP.

Similarly, for the second major party Indian National Congress, in Fig. 4, we can see that when we analysed the tweets without handling sarcasm, we got 6.50% of the total tweets to be of positive polarity, while 4.88% of the total tweets were of negative polarity. But when we handled the sarcasm using the transfer learning technique, then 6.34% of the total tweets were of positive polarity, while 6.64% of the tweets were of negative polarity for Indian National Congress (INC) as shown in Fig. 7. The above observations are represented in Table 3.

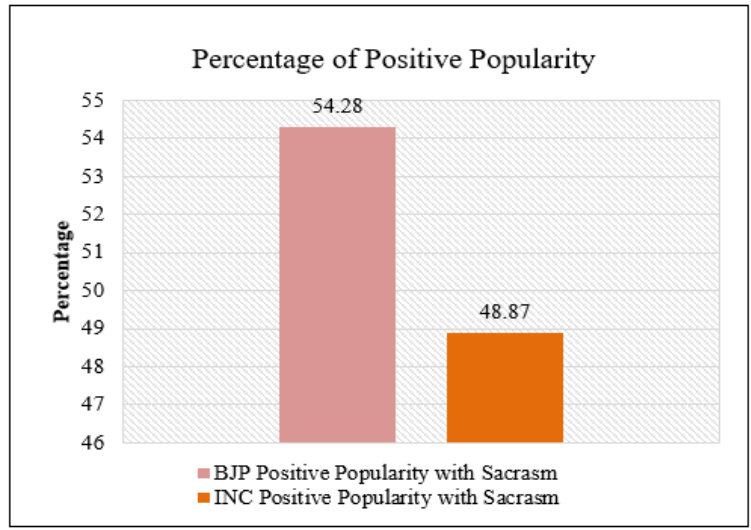

**Fig. 9.** Percentage of Positive tweets among total tweets with sarcasm handling

**Table 3.** Polarity Percentage of tweets with respect to total tweets

| Party | Positive Polarity of tweets (without sarcasm handling) | Positive Polarity of tweets (with sarcasm handling) | Negative Polarity of tweets (without sarcasm handling) | Negative Polarity of tweets (with sarcasm handling) |
|-------|---------------------------|---------------------------|---------------------------|---------------------------|
| BJP | 25.58% | 23.76% | 13.16% | 20.01% |
| INC | 6.50% | 6.34% | 4.88% | 6.64% |

To better understand the results, we plotted the bar graph for the positive to negative tweets for both political parties. Both the graphs for the cases without sarcasm handling and with sarcasm handling are shown in Fig. 5 and Fig. 8, respectively. The ratio of positive to negative tweets for the BJP party is 1.94, while that of INC is 1.33 without sarcasm handling, but when we handled the sarcasm using our trained model, the results changed for both cases. It became 1.19 for the BJP party, while for the INC party also dropped to 0.96. All the result values for the ratios can be represented using Table 4.

**Table 4.** Positive: Negative polarity ratio of tweets

| Party | Positive: Negative (without Sarcasm Handling) | Positive: Negative (with Sarcasm Handling) |
|-------|-----------------------------------------------|--------------------------------------------|
| BJP   | 1.94                                          | 1.19                                       |
| INC   | 1.33                                          | 0.96                                       |

Then, we generated another bar graph for the Percentage of positive tweets for a political party, with respect to the total tweets about that particular political party. The graphs for this value are represented in Fig. 6 (without sarcasm handling) and Fig. 9 (with sarcasm handling). We found in our prediction that 66.03% of the total tweets about the BJP party and its leaders were positive polarity and 57.14% of the total tweets about the INC party were positive polarity when sarcasm handling is not considered. But, when we tried to handle the sarcasm, we got the results that 54.28% of the total tweets about the BJP party were of positive polarity while 48.87% of the total tweets about the INC party were of negative polarity. The above results are summarized in Table 5.

**Table 5.** Percentage of positive tweets for a political party

| Party | Percentage (without Sarcasm handling) | Percentage (with sarcasm handling) |
|-------|---------------------------------------|------------------------------------|
| BJP   | 66.03%                                | 54.28%                             |
| INC   | 57.14%                                | 48.87%                             |

## 5      Conclusion and Future Works

This paper proposed an idea of transfer learning for handling sarcastic tweets and analysing the tweets for positive and negative polarities. It is observed that our trained models are working very well. When we compared our model's results with the actual election results, about 37.4% of the votes of the whole country were in the favour of the BJP party, while around 19.5% of the total votes were in favour of the INC party. Our model also predicted that the BJP party wins the election with a vote share difference of around 19% which is approximately similar to elections 2019 results.

The vote percentage difference in our model's results and the then election results are almost the same. Although the actual vote percentage for any party is having a difference with our model's predictions, the reasons behind this difference may be the following points, which need to be considered for a diversified country like India. There is a digital divide problem, which means that not an entire population has access to the Internet. Even among the people having access to the Internet, not everyone uses Twitter. Also, it could be the case that there could be some discrepancies while collecting the dataset. Further, in a society, many people avoid sharing negative emotions on public platforms due to many political and personal reasons.

In the future, the work can be extended to include the following. It can identify the concerned category for any negative tweet, like Agriculture, Education, Infrastructure, Price hike, which will be useful for the country's development. It can filter out the hate speech and abusive content related tweets, such that only constructive criticism related tweets will be considered. Further, to improve the model, instead of the transfer learning approach, some other more efficient unsupervised learning algorithm can be applied for upgrading the model.